\documentclass[aps,prl,superscriptaddress,amsmath,amssymb,floatfix,twocolumn,showpacs]{revtex4}

\pdfoutput=1

\usepackage{amsmath, dcolumn, bm, amsthm, amsfonts, amssymb, tabularx, slashbox, subfigure}
\usepackage[colorlinks=true,urlcolor=blue,citecolor=blue,linkcolor=blue]{hyperref}
\usepackage{times,bbold}
\usepackage{graphicx,graphics,color}

\begin{document}

\title{Inhomogeneous Topological Superfluidity in One-Dimensional Spin-Orbit-Coupled Fermi Gases}

\author{Chun Chen}
\email[]{cchen@physics.umn.edu}
\affiliation{School of Physics and Astronomy, University of Minnesota, Minneapolis, Minnesota 55455, USA}

\begin{abstract}

We theoretically predict an exotic topological superfluid state with spatially modulated pairing gap in one-dimensional spin-orbit-coupled Fermi gases. This inhomogeneous topological superfluidity is induced by applying simultaneously a perpendicular Zeeman magnetic field and an equally weighted Rashba and Dresselhaus spin-orbit coupling in one-dimensional optical lattices. Based on the self-consistent Bogoliubov--de Gennes theory, we confirm that this novel topological phase is a unique condensation of Cooper pairs, which manifests the interplay between the inhomogeneity of superfluid and its nontrivial topological structure. The properties of the emergent Majorana bound states are investigated in detail by examining the associated $\mathbb{Z}_{2}$ topological number, the eigenenergy and density of states spectra, as well as the wave functions of the localized Majorana end modes. Experimental feasibility of observing this new topological state of matter is also discussed.

\end{abstract}


\maketitle

{\it Introduction.}---The recent pursuit of Majorana fermions in nanomaterials, solid state systems, and atomic Fermi gases has generated much interest in both fields of condensed matter and cold atom physics \cite{majorana,wilczek,hasan,xlqi,alicea1,read,kitaev1}. A series of intriguing heterostructures comprising $s$-wave superconductors, topological insulators, ordinary semiconductors, as well as other ferromagnetic substances has been proposed to possess the capacity for harboring non-Abelian Majorana zero modes at interfaces of the sample through a proximity effect \cite{fu,sau1,lutchyn,oreg,alicea2,sau2}. Experimental realizations of these proposals have reported detection of zero-bias mid-gap states in InSb nanowires contacted with the normal metal and superconducting electrodes \cite{mourik,das,deng}.

A parallel extensive search for the Majorana bound states has also been progressing in the systems of ultracold fermionic superfluids \cite{cwzhang,sato1,sato2,jiang,wei,xjliu,fwu}. Particularly, the breakthrough of realizing synthetic gauge fields in cold atom condensates \cite{yjlin,pjwang,cheuk,galitski} greatly stimulates the research of topological superfluidity in spin-orbit-coupled (SOC) Fermi gases subject to the Zeeman field. One peculiar advantage of deploying ultracold atoms to probe the topological properties of a quantum fluid lies in the fact that via standard techniques of optical lattices and Feshbach resonances \cite{blochrev,chin}, now we can not only precisely tune the inter-particle interactions over a wide range of parameters, but also have the freedom to switch the dimension, modify the geometry, and control the purity of a quantum gas to experimentally simulate various theoretical models in modern physics \cite{bloch3}. In comparison with the well-studied two-dimensional structures, one-dimensional ($1$D) nanowires and optical lattices have recently attracted growing attention in detecting and engineering Majorana zero modes due to their simple $1$D confinement geometry and the resultant reduction of decoherence effects \cite{potter1,potter2,jiang,wei,xjliu,franz}. To some extent, fermionic cold atoms trapped in a tube can be free of disorders, and provide another ideal platform to further explore the new topological states of quantum matter.

In this Letter, we try such a new route to theoretically predict a topologically nontrivial Fulde-Ferrell-Larkin-Ovchinnikov (\emph{topo}-FFLO) superfluid state \cite{fulde,larkin,xjliu2,radzihovsky} in the $1$D SOC Fermi gases \cite{liao,pjwang,cheuk}. Given that high purity and strong superfluid pairing are both achievable in atomic systems \cite{blochrev,chin}, the proposed \emph{topo}-FFLO phase will offer a new setting for the advancement of understanding the relation and interplay between the inhomogeneity of superfluid and its intrinsic topological properties. As a new quantum state of matter exhibiting superfluidity, inhomogeneity, and Majorana fermions as a whole, this unified \emph{topo}-FFLO state may potentially suggest a promising strategy to study the superfluid inhomogeneity, its nontrivial topology, and the artificial disorders on the same footing in a highly-controllable cold atom laboratory \cite{jli}.

{\em Model and phase diagrams.}---Here we conceive a minimal $1$D lattice model, which simultaneously hosts Majorana fermions at the edges and an inhomogeneous FFLO phase in the bulk, to demonstrate the existence of a novel inhomogeneous topological superfluid in fermionic condensates. The Hamiltonian describing the $1$D SOC Fermi gases can be written as \cite{sato1,alicea1,alicea2,wei,xjliu,fwu}:
\begin{eqnarray}
H\!\!&=&\!\!H_{\textrm{K}}+H_{\textrm{R}}+H_{\textrm{D}}+H_{\Delta}, \nonumber \\
H_{\textrm{K}}\!\!&=&\!\!-t \!\sum_{ i,j, \sigma }\! \psi^{\dagger}_{i\sigma} \psi_{j\sigma}\! + \!\sum_{i, \sigma} \mbox{[} V \mbox{(}r_{i}\mbox{)} +h\sigma_{z}-\mu \mbox{]} \psi^{\dagger}_{i\sigma} \psi_{i\sigma},  \nonumber \\
H_{\textrm{R}}\!\!&=&\!\! -\lambda_{z}\! \sum_{i} \mbox{(}\psi^{\dagger}_{i\downarrow} \psi_{i+\hat{x}\uparrow}-\psi^{\dagger}_{i\uparrow} \psi_{i+\hat{x}\downarrow}+\textrm{H.c.}\mbox{)}, \label{eq:eqn1_rs} \\
H_{\textrm{D}}\!\!&=&\!\! -\lambda_{y} \!\sum_{i} \mbox{(}i\psi^{\dagger}_{i\uparrow} \psi_{i+\hat{x}\uparrow}-i\psi^{\dagger}_{i\downarrow} \psi_{i+\hat{x}\downarrow}+\textrm{H.c.}\mbox{)}, \nonumber \\
H_{\Delta}\!\!&=&\!\! -\!\sum_{i}\mbox{(}\Delta_{i} \psi^{\dagger}_{i\uparrow} \psi^{\dagger}_{i\downarrow} +\Delta^{\ast}_{i} \psi_{i\downarrow} \psi_{i\uparrow}\mbox{)}, \nonumber
\end{eqnarray}
where $\psi^{\dagger}_{i\sigma}$ ($\psi_{i\sigma}$) denotes the creation (annihilation) field operator with spin $\sigma\equiv \mbox{(} \! \uparrow \! \mbox{,} \! \downarrow \! \mbox{)}$ at site $r_{i}$. $H_{\textrm{K}}$ is the kinetic term including the nearest neighbor hopping $t$, the $1$D harmonic trapping $V\mbox{(}r_{i}\mbox{)} \! \equiv \!m \omega^{2}r^{2}_{i}/2$, a perpendicular Zeeman field $h$, and chemical potential $\mu$. In consideration of the current experimental status that only $1$D equally weighted Rashba and Dresselhaus (ERD) spin-orbit (SO) coupling has been realizable in cold gases of fermions, we explicitly adopt two different kinds of SO interactions for a general purpose \cite{alicea2,supplemental}. $H_{\textrm{R}}$ represents the spin-flip Rashba type SO coupling with strength $\lambda_{z}$ \cite{rashba}, while $H_{\textrm{D}}$ is a spin-conserving Dresselhaus ($110$) SO interaction with strength $\lambda_{y}$ \cite{dresselhaus}. $H_{\Delta}$ denotes the spin-singlet $s$-wave contact attraction between atoms with the gap function $\Delta_{i}\!\equiv\! V_{s} \langle \psi_{i\downarrow} \psi_{i\uparrow} \rangle$, whence $V_{s}$ is the pairing strength.

To visualize the landscape of candidate ground states, we first focus on understanding the bulk properties of the $1$D SOC Fermi gas by ignoring the trapping potential and imposing the periodic boundary condition. The resulting Hamiltonian in the momentum space can be expressed via Nambu spinor $\Psi_{k}^{\dagger}\equiv\mbox{(}\psi^{\dagger}_{\frac{q}{2}+k \uparrow}~~\psi^{\dagger}_{\frac{q}{2}+k \downarrow}~~\psi_{\frac{q}{2}-k \uparrow}~~\psi_{\frac{q}{2}-k \downarrow}\mbox{)}$ as $H=\sum_{k} \Psi_{k}^{\dagger} \mathcal{H}\mbox{(}k\mbox{)} \Psi_{k}$ up to a constant, where the Bogoliubov--de Gennes (BdG) Hamiltonian reads:
\begin{eqnarray}
\mathcal{H}\mbox{(}k\mbox{)}\!\!&=&\!\!\frac{1}{2}\!
    \left(\begin{array}{cccc}
    \xi^{+}_{\frac{q}{2}+k} & \eta_{\frac{q}{2}+k} & 0 & -\Delta_{q} \\
    -\eta_{\frac{q}{2}+k} & \xi^{-}_{\frac{q}{2}+k} & \Delta_{q} & 0 \\
    0 & \Delta_{q} & -\xi^{+}_{\frac{q}{2}-k} & \eta_{\frac{q}{2}-k} \\
    -\Delta_{q} & 0 & -\eta_{\frac{q}{2}-k} & -\xi^{-}_{\frac{q}{2}-k} \\
    \end{array}\right)\!\!. \label{eq:eqn2_ks}
\end{eqnarray}
In Eq.~(\ref{eq:eqn2_ks}), we assume the superfluid order is composed of the pair condensation at a specific center-of-mass (COM) momentum $q$: $\Delta_{i}\!=\!\Delta_{q} e^{i q r_{i}}$ \cite{xjliu2,radzihovsky}, and dispersions $\xi^{\pm}_{\frac{q}{2}\pm k}\!=\!-2t\cos\mbox{(}\frac{q}{2}\pm k\mbox{)}-\mu\pm \mbox{[}h+2\lambda_{y}\sin\mbox{(}\frac{q}{2}\pm k\mbox{)}\mbox{]}$; $\eta_{\frac{q}{2}\pm k}\!=\!2i\lambda_{z}\sin\mbox{(}\frac{q}{2}\pm k\mbox{)}$. Typically, the eigenenergy of $\mathcal{H}\mbox{(}k\mbox{)}$ cannot be analytically manipulated, so we need solve the problem numerically and minimize mean-field thermodynamic potential $E_{g}\!=\!\langle H \rangle /N\! +\! \Delta^{2}_{q}/V_{s}$ at zero temperature to self-consistently extract the values of $q$ and $\Delta_{q}$ for the ground state.

Moreover, by noticing that $\mathcal{H}\mbox{(}k\mbox{)}$ respects the built-in particle-hole symmetry, we can thus introduce an associated $1$D $\mathbb{Z}_{2}$ number $(-1)^{\nu}$, where the Berry phase $\nu$ equals $\frac{i}{\pi}\sum_{E\mbox{(}k\mbox{)}<0}\int^{\pi}_{-\pi} \langle\phi\mbox{(}k\mbox{)}|\partial_{k}\phi\mbox{(}k\mbox{)}\rangle dk$, to characterize the nontrivial topological structure of the Bloch bands in the presence of both SO interactions and Zeeman field \cite{supplemental}. When the $\mathbb{Z}_{2}$ number is $-1$ ($+1$), the bulk system will be topologically nontrivial (trivial). It is worth mentioning that to engender a topological superfluid, besides the nontrivial band structure, the system also needs to sustain the channel of an effective spinless $p$-wave pairing. In our model [Eq.~(\ref{eq:eqn2_ks})], a nonzero $\Delta_{q}$ will just encode the designed $p$-wave symmetry for the intraband pairing once projected into the helical basis. Therefore, it becomes practical to employ the pair condensation and the $1$D $\mathbb{Z}_{2}$ topological invariant as a composite order parameter to discriminate among the differing ground states in general and accordingly map out the phase diagrams of the Fermi gases.

\begin{figure}
\begin{center}
\includegraphics[width=7.6cm]{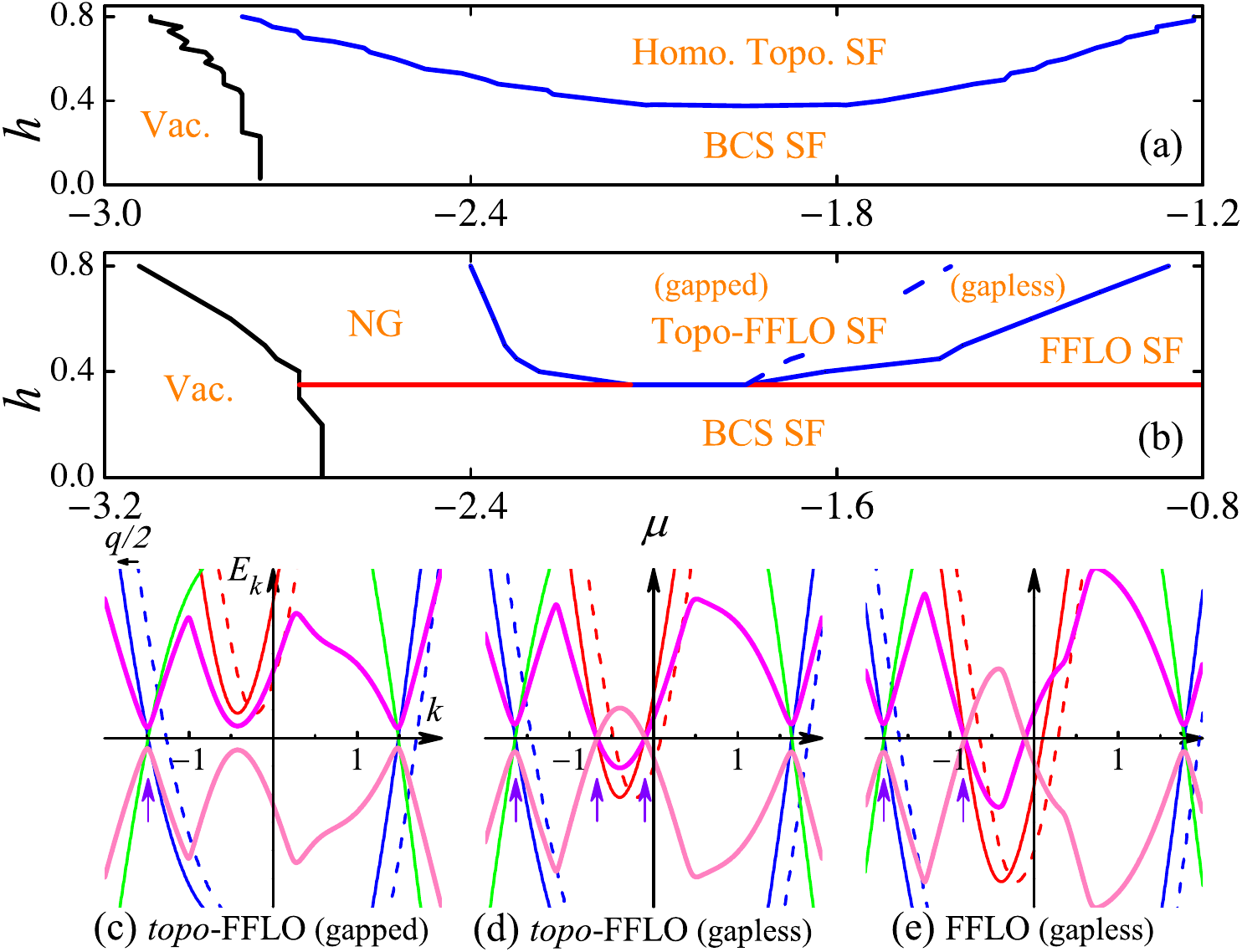}
\caption{\label{fig:fig1} (color online). Two generic phase diagrams of a $1$D SOC fermionic condensate on the $h$-$\mu$ coordinates. Panel (a) is for homogeneous superfluid (SF) states with Rashba SO coupling ($\lambda_{z}\!=\!\lambda$, $\lambda_{y}\!=\!0$), while panel (b) illustrates the emergence of an inhomogeneous topological superfluid (gapped and gapless) in the ERD SOC Fermi gas ($\lambda_{z}\!=\!\lambda_{y}\!=\!\frac{\sqrt{2}}{2}\lambda$) [$\lambda\!=\!0.7t$]. The phase boundaries are symmetric to $\mu\!=\!0$ and NG stands for the normal gas state. Three lower panels (c)-(e) show the dispersion spectra of varied FFLO states in (b) with fixed $h\!=\!0.6t$ and increasing $\mu\!=\!-1.6t~\mbox{(c)}$; $-1.3t~\mbox{(d)}$; $-1.0t~\mbox{(e)}$, where thin solid (dashed) lines [red and blue] represent (un)shifted helical bands of the non-interacting system. The thick magenta line denotes the lowest quasihole branch of Eq.~(\ref{eq:eqn2_ks}). Purple arrows mark the Fermi points of shifted non-interacting helicity bands in half of the Brillouin zone (BZ).}
\end{center}
\end{figure}

In all self-consistent calculations, we set $t\!=\!1$ as the energy unit, the $s$-wave attraction magnitude $V_{s}\!=\!2.5t$, and the strengths of Rashba and Dresselhaus SO couplings are determined by $\lambda^{2}_{z}\!+\!\lambda^{2}_{y}\!=\!\lambda^{2}$ [$\lambda\!=\!0.7t$]. For simplicity, we only present the results at zero temperature. Figure~$\ref{fig:fig1}$ shows the typical phase diagrams for the $1$D Fermi gases at different intensities of SO couplings on the $h$-$\mu$ plane. In panel (a), we keep $\lambda_{z}\!=\!\lambda$ and the resulting ground states in the limit of strong Rashba interaction are thus dominated by the homogeneous states with vanishing $q$. Specifically, we observe that besides vacuum (empty bands without filling) and conventional Bardeen-Cooper-Schrieffer (BCS) superfluid, there exists another homogeneous but topologically nontrivial superfluid phase in the region of large $h$. All these results are consistent with previous studies on the Rashba SOC Fermi gases \cite{wei,xjliu}. The phase diagram will get much richer and more interesting if we turn on the strong ERD SO interaction [Fig.~$\ref{fig:fig1}$(b)]. Instead of the homogeneous phases, we find that under strong Zeeman fields ($h\!>\!0.35t$), the whole system would be driven from a BCS superfluid into an inhomogeneous pairing state with a finite COM momentum. Through tuning the band-filling, we can further get access to the topological portion of the spectrum, hence triggering the emergence of a new type of \emph{topo}-FFLO superfluid carrying simultaneously a finite $q$, a nonzero $\Delta_{q}$, and a nontrivial $\mathbb{Z}_{2}$ number \lq\lq $-1$''. By examining quasiparticle excitations, we can also distinguish between gapped and gapless \emph{topo}-FFLO states. The physical underpinning of this inhomogeneous topological superfluidity is mainly stemming from the inversion \emph{asymmetry} of Bloch bands, which means that with fixed ERD SO coupling, the increase of $h$ will not only modify the topology of the band structure via opening a spin-orbit gap and spoiling time reversal symmetry, but it also facilitates the effective $p$-wave fermionic pairing at a nonvanishing COM momentum $q$. It is even appealing to perceive that the inclusion of Dresselhaus SO coupling would also efficiently \emph{enlarge} the domain of FFLO state in the phase diagram.

To clarify the origin of various FFLO states in Fig.~$\ref{fig:fig1}$(b), we highlight here their dispersions near the Fermi level. Figures~$\ref{fig:fig1}$ [(c)-(e)] show the following: (i) Due to the inversion asymmetry, superfluid pairing \emph{inside} the lower \emph{spin-mixed} helicity branch (blue solid line) opens the energy gaps at its Fermi points and shifts the whole bands to the left side by $q/2$. This also gives rise to the gapless state if $\mu$ crosses the upper helical band (red solid line). (ii) The $\mathbb{Z}_{2}$ topological invariant can now be approximate to the parity of the number of Fermi points in half of the BZ (marked by purple arrows). (iii) Physically separated edge modes may appear in gapped \emph{topo}-FFLO phase, namely when Fermi level lies within the spin-orbit gap.

\begin{figure}
\begin{center}
\includegraphics[width=7.6cm]{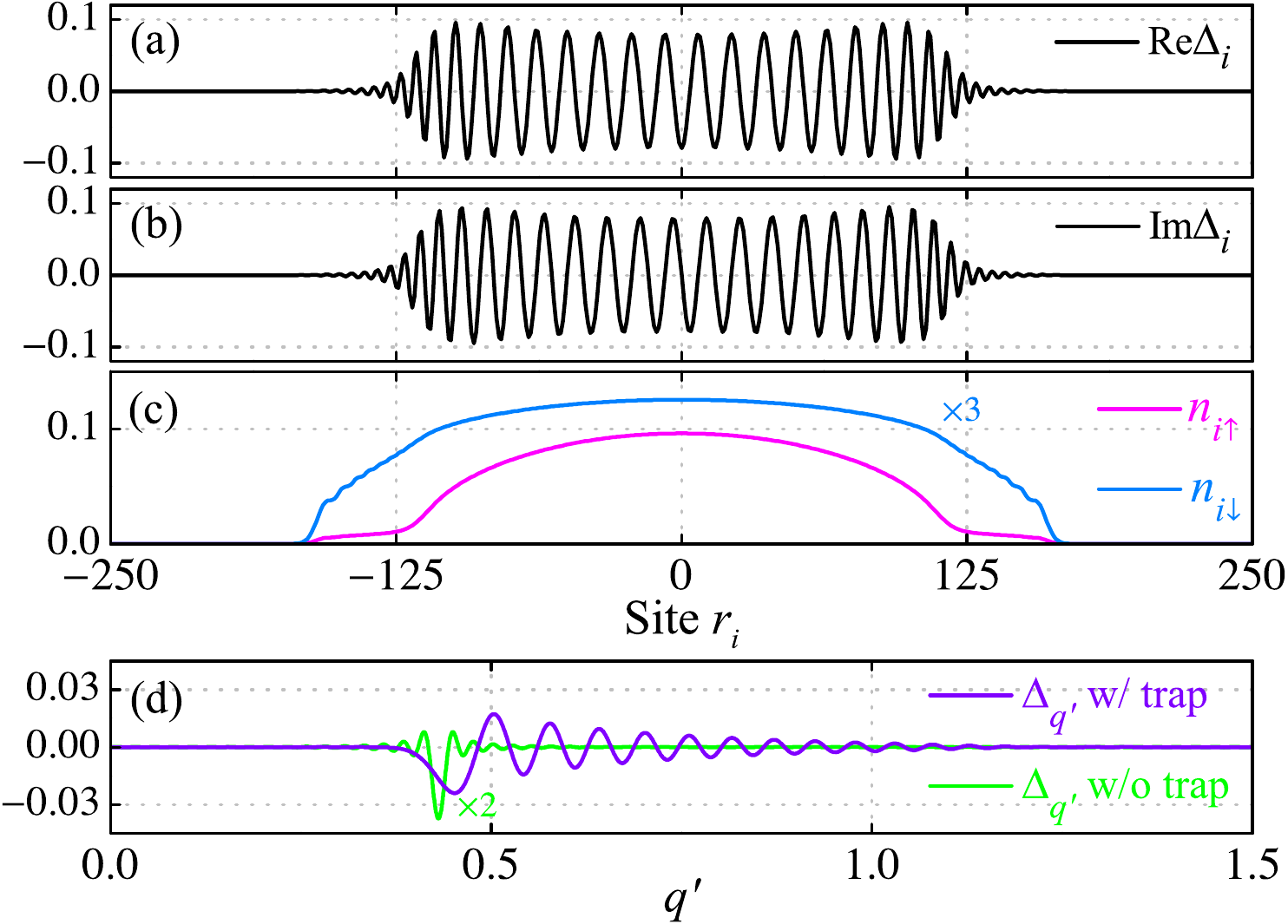}
\caption{\label{fig:fig2} (color online). Spatial profiles of superfluid order parameter $\Delta_{i}$ and atomic densities $n_{i\sigma}$ for the $1$D ERD SOC Fermi gas in real-space confinement. The real and imaginary parts of $\Delta_{i}$ are plotted in panels (a) and (b), respectively. Density distributions $n_{i\sigma}$ are shown in panel (c). Fourier transforms of $\Delta_{i}$ at finite momenta $q'$ are depicted in panel (d), where the purple (green) line denotes $\Delta_{q'}$ with (without) trap. Here $V_{s}\!=\!2.5t$, $\lambda\!=\!0.7t$, $h\!=\!0.6t$, $\mu\!=\!-1.6t$, $\lambda_{z}\!=\!\lambda_{y}\!=\!\frac{\sqrt{2}}{2}\lambda\!\approx\!0.495t$, and $m\omega^{2}/2\!\approx\!0.0001t$.}
\end{center}
\end{figure}

{\em Topo-FFLO superfluid in real space.}---Now we concentrate on real-space configurations of the emergent Majorana end states in \emph{topo}-FFLO phase (gapped) under realistic harmonic trapping and ERD SO coupling. After performing canonical transformations: $\psi_{i\uparrow}\!=\!\sum_{n}\mbox{[}u_{n\uparrow}\mbox{(}r_{i}\mbox{)} \gamma_{n\uparrow}\!-\!v^{\ast}_{n\downarrow}\mbox{(}r_{i}\mbox{)} \gamma^{\dagger}_{n\downarrow}\mbox{]}$ and $\psi^{\dagger}_{i\downarrow}\!=\!\sum_{n}\mbox{[}v_{n\uparrow}\mbox{(}r_{i}\mbox{)} \gamma_{n\uparrow}\!-\!u^{\ast}_{n\downarrow}\mbox{(}r_{i}\mbox{)} \gamma^{\dagger}_{n\downarrow}\mbox{]}$, we can obtain the self-consistency BdG equations from Eq.~(\ref{eq:eqn1_rs}) as follows: $\mbox{[}H,\gamma_{n\sigma}\mbox{]}\!=\!-E_{n}\gamma_{n\sigma}$ and $\mbox{[}H,\gamma^{\dagger}_{n\sigma}\mbox{]}\!=\!E_{n}\gamma^{\dagger}_{n\sigma}$. Since the system preserves particle-hole symmetry, Bogoliubov quasiparticle operators $\gamma$'s satisfy the relations $\gamma_{-E}\!=\!\gamma^{\dagger}_{E}$, which implies that the system will become topologically nontrivial if $E\!=\!0$ and $\gamma_{0}\!=\!\gamma^{\dagger}_{0}$. The real-space computation [Fig.~$\ref{fig:fig2}$] is conducted on a $501\!\times\!1$ lattice with an open boundary condition [$V_{s}\!=\!2.5t$, $h\!=\!0.6t$, $\mu\!=\!-1.6t$, $\lambda_{z}\!=\!\lambda_{y}\!=\!\frac{\sqrt{2}}{2}\lambda$, and $m\omega^{2}/2\!\approx\!0.0001t$].

Figure~$\ref{fig:fig2}$ summarizes the spatial profiles of superfluid order parameter and fermion density distributions in the confined $1$D tube along $\hat{x}$-direction. Driven by the interplay between the asymmetry of Fermi surface (points) and the superfluid pairing, real and imaginary parts of $\Delta_{i}$ display rapid oscillations across the zero point in antisymmetric ways throughout the whole region of the quantum gas with a period of about $12$ sites [see Figs.~$\ref{fig:fig2}$(a) and (b)]. To gain a concrete understanding of this spatial variation, we expand $\Delta_{i}$ in terms of a spectrum of plane waves $\Delta_{i}\!=\!\sum_{q'} \Delta_{q'} e^{i q' r_{i}}$. It becomes clear from Fig.~$\ref{fig:fig2}$(d) that the imposition of a harmonic trap induces the modulation of multiple Fourier modes $\Delta_{q'}$ in the interval ranging from $q'\!\sim\!0.5$ to $1.0$. While, for the case of free gas without trap, $\Delta_{q'}$ change to centrally distribute around a single momentum $q'\!\approx\!0.45$ [see the green line in Fig.~$\ref{fig:fig2}$(d)]. This is consistent with our $k$-space formalism. With such observations, we deduce that at the given parameters, the bulk system has entered an inhomogeneous FFLO state. In Fig.~$\ref{fig:fig2}$(c) we add the density distributions $n_{i\sigma}$ as well, from which the bimodal structure of spin $\uparrow$ atom distribution is visible.

\begin{figure}
\begin{center}
\includegraphics[width=7.6cm]{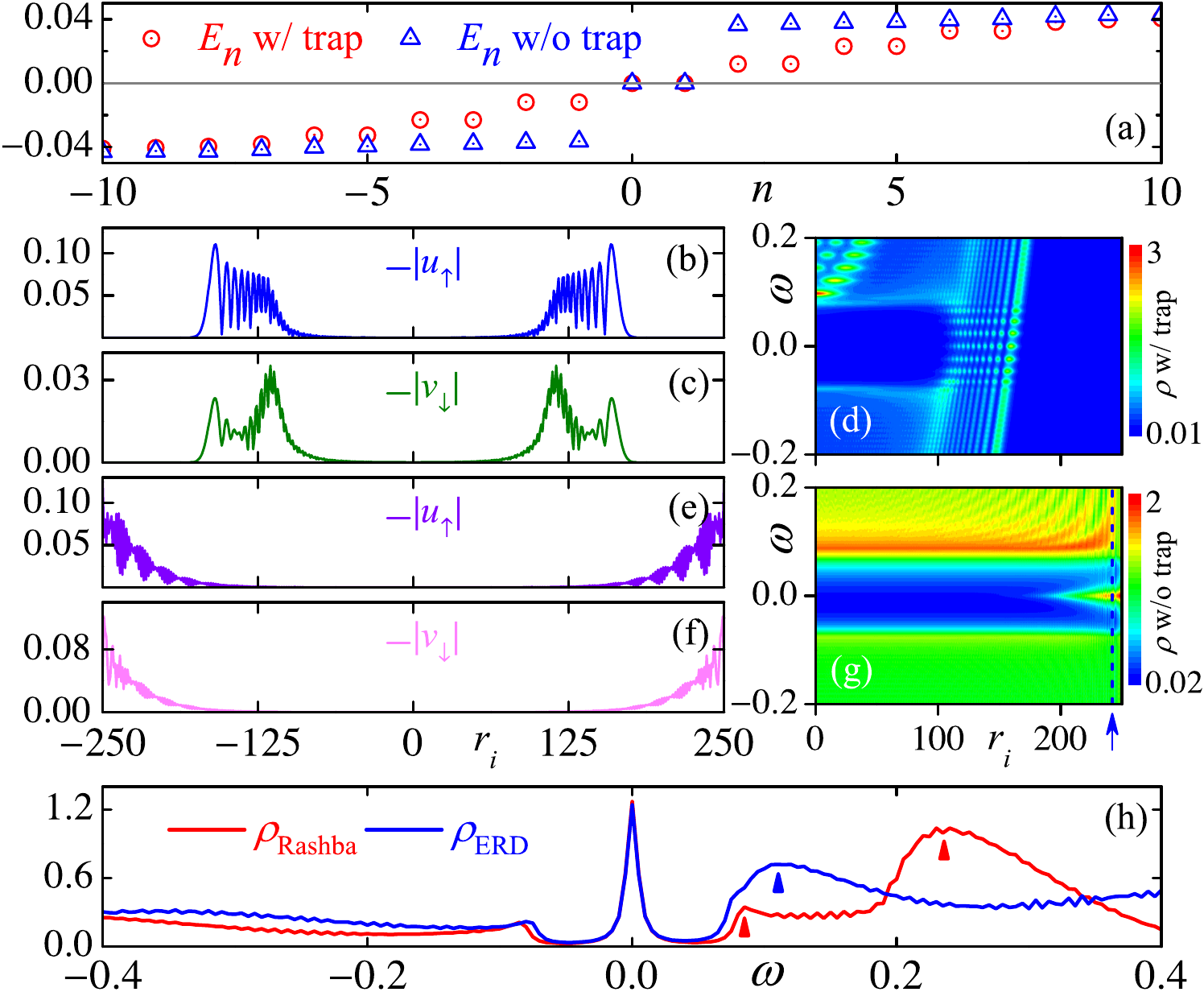}
\caption{\label{fig:fig3} (color online). Quasiparticle spectra $E_{n}$ of the $1$D ERD SOC Fermi gases with (red circle) and without (blue triangle) trapping $\mbox{[panel (}\textrm{a}\mbox{)]}$. Panels [(b),(c)] and [(e),(f)] depict the amplitudes of wave functions for the zero energy states with and without trap, respectively. Panels (d) (with trap) and (g) (without trap) show the corresponding contours of LDOS along half of the lattice. All parameters are the same as in Fig.~$\ref{fig:fig2}$. Total LDOS spectra along the cut at site $r_i=242$ [see blue dashed line in (g)] without trap are plotted in panel (h) for both topological BCS superfluid [$\rho_{\textrm{Rashba}}$; $\lambda_{z}\!=\!\lambda$, $\lambda_{y}\!=\!0$] and \emph{topo}-FFLO superfluid [$\rho_{\textrm{ERD}}$; $\lambda_{z}\!=\!\lambda_{y}\!=\!\frac{\sqrt{2}}{2}\lambda\!\approx\!0.495t$]. Here $\rho_{i}\mbox{(}\omega\mbox{)}=\sum_{n,\sigma}\mbox{[}|u_{n\sigma}\mbox{(}r_{i}\mbox{)}|^2\delta\mbox{(}E_{n}-\omega\mbox{)}+|v_{n\bar{\sigma}}\mbox{(}r_{i}\mbox{)}|^2\delta\mbox{(}E_{n}+\omega\mbox{)}\mbox{]}$.}
\end{center}
\end{figure}

Remarkably, when mapping out the corresponding excitation spectrum of the system, we find that interestingly, the $1$D quantum gas inside tube also possesses nontrivial topological properties. As shown in Fig.~$\ref{fig:fig3}$(a), after switching on ERD SO interaction, the minimal value of eigenenergy $|E_{n}|$ becomes exponentially small ($\sim\!10^{-10}$), which indicates the emergence of unpaired Majorana fermions at the edges. Note that there is only \emph{one} pair of Majorana fermions survived in our system. The second lowest value of $|E_{n}|$ equals $0.012$ with trap and $0.036$ without trap. Since these gapless chiral edge states live only inside domain walls separating topologically distinct regimes, this remaining pair of Majorana zero modes is signalling that besides vacuum and normal gas state, there should just exist one united quantum phase formed by fermionic atoms in trapped $1$D optical lattice. This unique condensation of Cooper pairs is exactly the \emph{topo}-FFLO state we discussed. In Figs.~$\ref{fig:fig3}$ [(b),(c);(d)] (with trap) and [(e),(f);(g)] (without trap), we further plot separately the amplitudes of wave functions and the corresponding local density of states (LDOS) spectra for the unpaired Majorana fermions, from which we can see that the self-conjugate zero modes localized at boundaries resemble Jackiw-Rebbi solution in Dirac equation \cite{supplemental,jackiw}. One distinguishing feature of the LDOS contours is the presence of mid-gap zero-bias peak at trap edge, which serves as a compelling evidence for the realization of Majorana fermions in $1$D SOC chain. Moreover, being specific to \emph{topo}-FFLO phase, in Fig.~$\ref{fig:fig3}$(h), we find that its spectral weights near positive gap edge get a dramatic enhancement as compared to the LDOS of topological BCS superfluid (marked by triangles). This measurable feature may help differentiate these two topological states in experiments.

Similar to the mechanism proposed for semiconducting heterostructures \cite{alicea1,alicea2}, this revealed coexistence of topological order and FFLO superfluidity is yielded by the conspiracy of a spin-singlet--pairing mediated $p$-wave superfluid instability in the topologically nontrivial Bloch bands with the Zeeman field facilitated breaking of time reversal and \emph{inversion} symmetries. Non-Abelian Majorana fermions are then emerging from the phase twist of orbital motion accompanying ERD SO interaction \cite{sato1,sato2}. In our perspective, the lattice model Eq.~(\ref{eq:eqn1_rs}) demonstrates the first attainable scenario of creating the predicted inhomogeneous topological superfluidity in atomic Fermi gases \cite{yjlin,pjwang,cheuk,liao}. It also uncloaks a novel mechanism for the FFLO superfluidity/superconductivity in a single spin-mixed asymmetric helicity band of a SOC system.

{\em Experimental realization.}---We propose to use fermionic lithium atoms as a quantum simulator to synthesize and detect this new topological state of matter in cold atom condensates. It has been shown that spin-imbalanced $^{6}$Li degenerate gas loaded in an array of tubes can realize the partially polarized superfluid phase with possible FFLO correlations in $1$D \cite{liao}. The $^{6}$Li atoms can also be dressed up via a pair of Raman beams to produce the $1$D equal-part Rashba and Dresselhaus SO coupling, and the opened spin-orbit gap has been directly observed in spin-injection spectroscopy measurements \cite{cheuk,galitski}. Therefore, all the required techniques of simulating the model Hamiltonian Eq.~(\ref{eq:eqn1_rs}) are within the scope of current experimental sophistication. Furthermore, signatures of \emph{topo}-FFLO state may be detectable with spatially resolved radio-frequency spectroscopy and time-of-flight imaging through seeking the described features in LDOS spectrum and the bimodal structure of atom density distribution \cite{supplemental}.

{\em Conclusions.}---As the first endeavor to predict a new quantum state of matter, our work provides the theoretical framework and presents the basic information of \emph{topo}-FFLO phase within mean-field theory and proper topological arguments. In view of its conceptual novelty and experimental feasibility, a more accurate description of this topological state necessitates the usage of advanced techniques, which will be left for future investigations \cite{mizushima,yang1,yang2,stoudenmire,mcheng,fidkowski}.

In summary, we theoretically study the phase diagram of an ERD SOC Fermi gas in $1$D optical lattice, and successfully identify \emph{for the first time} an exotic topological FFLO superfluid in the region of strong SO couplings and Zeeman field. Detailed structures of order parameters and Majorana end modes associated with this \emph{topo}-FFLO phase are manifestly uncovered in real space through utilizing the BdG formulation. Our work might open up new prospects for the exploration of topological states of matter in SOC systems.

C.C. thanks C.S. Ting, F.J. Burnell, A. Kamenev, and Y. Chen for discussions. This work was supported in part by Summer Research Fellowships at University of Minnesota.

{\em Note added.}---After completion of present paper, we learnt of three related works that addressed and confirmed the same phase in $2$D \cite{clqu,wzhang} and $1$D \cite{xjliu3} SOC Fermi gas systems.

\end{document}


\title{Supplemental Material}

\author{Chun Chen}
\email[]{cchen@physics.umn.edu}
\affiliation{School of Physics and Astronomy, University of Minnesota, Minneapolis, Minnesota 55455, USA}

\maketitle

In this Supplemental Material, we provide some detailed derivations and analyses on the form of the model Hamiltonian, the equivalent formulations of calculating the topological index, some details of the topological phase transition, and the analytical solution of the zero energy bound state, which may help interested researchers acquire a better understanding of the main paper. Some relevant comments and discussions on the experimental detection of the \emph{topo}-FFLO state and the validity of the employed mean-field theory are also included.

\subsection{I. Equally weighted Rashba and Dresselhaus spin-orbit coupling}

Starting from the Dirac equation by an expansion of $v/c$, we can obtain the general three dimensional ($3$D) spin-orbit (SO) interaction in the SI units as follows:
\begin{equation}
H^{\textrm{3D}}_{\textrm{SO}}=-\frac{e\hbar}{4 m^{2}c^{2}} \vec{\bm{\sigma}} \!\cdot\! \left[ \vec{\mathbf{E}}\times \left( \vec{\mathbf{p}}-\frac{e}{c} \vec{\mathbf{A}} \right) \right],
\end{equation}
where $\vec{\bm{\sigma}}$ are the usual Pauli matrices, and $\vec{\mathbf{E}}$ is the electric field or the gradient of a scalar potential. Let's neglect the vector potential $\vec{\mathbf{A}}$, and expand the cross product into
\begin{eqnarray}
\vec{\mathbf{E}} \times \vec{\mathbf{p}}&=&
    \left|\begin{array}{ccc}
    \bm{i} & \bm{j} & \bm{k} \\
    E_x & E_y & E_z \\
    p_x & p_y & p_z \\
    \end{array}\right| = (E_y p_z-E_z p_y)\bm{i}+(E_z p_x-E_x p_z)\bm{j}+(E_x p_y-E_y p_x)\bm{k},
\end{eqnarray}
then
\begin{equation}
H^{\textrm{3D}}_{\textrm{SO}}=-\frac{e\hbar}{4 m^{2}c^{2}} \left[\sigma_x (E_y p_z-E_z p_y)+\sigma_y (E_z p_x-E_x p_z)+\sigma_z (E_x p_y-E_y p_x)\right].
\end{equation}
$1^{\circ}$ If we choose $\vec{\mathbf{E}}=(0,0,E_z)$, $H^{\textrm{3D}}_{\textrm{SO}}$ is the $2$D Rashba SO term \cite{rashba}: $H^{\textrm{2D}}_{\textrm{R}}=\frac{e\hbar E_z}{4 m^{2}c^{2}}\left(\sigma_x p_y-\sigma_y p_x \right)$. \\ $2^{\circ}$ If we confine the system to $1$D along $\hat{x}$ direction, namely $\vec{\mathbf{p}}=(p_x,0,0)$, we will get the $1$D SO terms used in our paper \cite{birk}: $H^{\textrm{1D}}_{\textrm{SO}}=-\frac{e\hbar}{4 m^{2}c^{2}} \left(\sigma_y E_z p_x -\sigma_z E_y p_x \right)$. This form can be understood as follows: in a $1$D SOC chain constructed from a $3$D optical lattice, the effect of SO interaction due to the confinement in the $\hat{z}$ direction is accounted for by a spin-flip hopping, while the effect of SO interaction due to the confinement in the $\hat{y}$ direction is taken into account by an imaginary spin-conserving hopping.

In $2$D solid state systems, the structure inversion asymmetry of a confinement potential and/or the bulk inversion asymmetry of a crystal can lead to the well-known Rashba and Dresselhaus SO couplings, respectively \cite{rashba,dresselhaus}. The Rashba SO interaction in a square lattice has the form
\begin{equation}
H^{\textrm{2D}}_{\textrm{R}}\!=\!-\lambda_{z}\left\{ \sum_{i} \left[ \psi^{\dagger}_{i\downarrow} \psi_{i+\hat{x}\uparrow}-\psi^{\dagger}_{i\uparrow} \psi_{i+\hat{x}\downarrow}+\textrm{H.c.} \right] \!+\!\sum_{i} \left[ i \left( \psi^{\dagger}_{i\downarrow} \psi_{i+\hat{y}\uparrow}+\psi^{\dagger}_{i\uparrow} \psi_{i+\hat{y}\downarrow} \right)+\textrm{H.c.} \right] \right\}.
\end{equation}
Now for a $1$D chain along $\hat{x}$ direction, the second term on the right-hand side can be dropped. We are thus left with the $1$D Rashba term $H^{\textrm{1D}}_{\textrm{R}}\!=\! -\lambda_{z}\! \sum_{i} \mbox{(}\psi^{\dagger}_{i\downarrow} \psi_{i+\hat{x}\uparrow}-\psi^{\dagger}_{i\uparrow} \psi_{i+\hat{x}\downarrow}+\textrm{H.c.}\mbox{)}$. This is exactly the third line of Equation (1) in the manuscript, which corresponds to the term $-\frac{e\hbar}{4 m^{2}c^{2}} \left(\sigma_y E_z p_x\right)$. The Dresselhaus (110) SO interaction \cite{you} is defined as
\begin{eqnarray}
H^{\textrm{(110)}}_{\textrm{D}}\!&=&\!-\lambda_{y} \sum_{i,s,s'} i\left(\sigma_z\right)_{ss'} \left(\psi^{\dagger}_{i-\hat{x},s}  \psi_{i,s'}-\psi^{\dagger}_{i+\hat{x},s}  \psi_{i,s'}\right) \nonumber \\
&=&\!-\lambda_{y} \sum_{i} \left(i\psi^{\dagger}_{i-\hat{x}\uparrow}  \psi_{i\uparrow}-i\psi^{\dagger}_{i-\hat{x}\downarrow}  \psi_{i\downarrow}-i\psi^{\dagger}_{i+\hat{x}\uparrow}  \psi_{i\uparrow}+i\psi^{\dagger}_{i+\hat{x}\downarrow}  \psi_{i\downarrow}\right) \nonumber \\
&=&\!-\lambda_{y} \sum_{i} \left(i\psi^{\dagger}_{i\uparrow} \psi_{i+\hat{x}\uparrow}-i\psi^{\dagger}_{i\downarrow} \psi_{i+\hat{x}\downarrow}+\textrm{H.c.}\right),
\end{eqnarray}
which is exactly the fourth line of Equation (1) in the manuscript, corresponding to the term $\frac{e\hbar}{4 m^{2}c^{2}} \left(\sigma_z E_y p_x \right)$. Therefore, the ERD SO coupling we adopt here is an equal combination of these two kinds of SO interactions in $1$D [$\lambda_{z}=\lambda_{y}$].

After doing a spin rotation or a local gauge transformation [$\sigma_{x}\rightarrow\sigma_{y},~\sigma_{y}\rightarrow\sigma_{z},~\sigma_{z}\rightarrow\sigma_{x}$] into an equivalent representation, our Hamiltonian [Equation (1) in the manuscript] can be mapped to the model including both the National Institute of Standards and Technology (NIST) SO coupling [$-\lambda \sigma_{y} p_{x}$] and the external Zeeman field $\vec{\mathbf{B}}=(0,B_{y},B_{z})$ \cite{yjlin,pjwang,cheuk}.

\subsection{II. Evaluation of the $\mathbb{Z}_{2}$ topological invariant}

In reference \cite{kitaev1}, Kitaev considered a periodic chain of $L$ unit cells, and each unit cell having $n$ fermionic sites. A general Hamiltonian for this $1$D model can be written in real space as [see Equations (18), (22), and (23) in ref.~\cite{kitaev1}]:
\begin{equation}
H=\frac{i}{4} \sum_{l,m} \sum_{\alpha,\beta} B_{\alpha \beta} (m-l) c_{l\alpha} c_{m\beta};~~~~B_{\alpha \beta} (j)^{\ast}=B_{\alpha \beta} (j)=-B_{\beta \alpha} (-j),
\end{equation}
where $l=1,\ldots,L$, $\alpha=1,\ldots,2n$, and the Majorana operators $c$'s satisfy the relations $c^{\dagger}_{m}=c_{m}$, $c_{l}c_{m}+c_{m}c_{l}=\delta_{lm}$. By performing the Fourier transformations, $\gamma_{p\alpha}=\frac{1}{\sqrt{L}}\sum_{j} e^{-ipj} c_{j\alpha}$, we can obtain the same Hamiltonian in the momentum space as,
\begin{equation}
H=\frac{i}{4} \sum_{\alpha,\beta} \sum_{p} \tilde{B}_{\alpha \beta} (p) \gamma_{-p\alpha} \gamma_{p\beta};~~~~\tilde{B}^{\dagger} (p)=-\tilde{B} (p)=\tilde{B}^{\textrm{T}} (-p),
\end{equation}
where $\tilde{B}_{\alpha \beta}(p)=\sum_{j} e^{ipj} B_{\alpha \beta}(j)$. Now let's turn our attention to the model studied in the manuscript [Equation (2)]. Via Nambu spinor $\Psi_{k}^{\dagger}\equiv(\psi^{\dagger}_{\frac{q}{2}+k \uparrow}~~\psi^{\dagger}_{\frac{q}{2}+k \downarrow}~~\psi_{\frac{q}{2}-k \uparrow}~~\psi_{\frac{q}{2}-k \downarrow})$, the Hamiltonian can be expressed as $H=\sum_{k} \Psi_{k}^{\dagger} \mathcal{H}(k) \Psi_{k}$ up to a constant, where the Bogoliubov--de Gennes (BdG) Hamiltonian reads
\begin{eqnarray}
\mathcal{H}(k)&=&\frac{1}{2}
    \left(\begin{array}{cccc}
    \xi^{+}_{\frac{q}{2}+k} & \eta_{\frac{q}{2}+k} & 0 & -\Delta_{q} \\
    -\eta_{\frac{q}{2}+k} & \xi^{-}_{\frac{q}{2}+k} & \Delta_{q} & 0 \\
    0 & \Delta_{q} & -\xi^{+}_{\frac{q}{2}-k} & \eta_{\frac{q}{2}-k} \\
    -\Delta_{q} & 0 & -\eta_{\frac{q}{2}-k} & -\xi^{-}_{\frac{q}{2}-k} \\
    \end{array}\right).
\end{eqnarray}
Changing to the Majorana representation $\Upsilon_{k}^{\dagger}\equiv(\gamma^{\textrm{A}}_{-k \uparrow}~~\gamma^{\textrm{B}}_{-k \uparrow}~~\gamma^{\textrm{A}}_{-k \downarrow}~~\gamma^{\textrm{B}}_{-k \downarrow})$ by the following transformation, $\Psi_{k}=\hat{\textrm{T}}\Upsilon_{k}$, namely
\begin{eqnarray}
\left(\begin{array}{c}
    \psi_{\frac{q}{2}+k \uparrow} \\
    \psi_{\frac{q}{2}+k \downarrow} \\
    \psi^{\dagger}_{\frac{q}{2}-k \uparrow} \\
    \psi^{\dagger}_{\frac{q}{2}-k \downarrow} \\
    \end{array}\right)&=&
    \left(\begin{array}{cccc}
    \frac{1}{\sqrt{2}} & \frac{i}{\sqrt{2}} & 0 & 0 \\
    0 & 0 & \frac{1}{\sqrt{2}} & \frac{i}{\sqrt{2}} \\
    \frac{1}{\sqrt{2}} & \frac{-i}{\sqrt{2}} & 0 & 0 \\
    0 & 0 & \frac{1}{\sqrt{2}} & \frac{-i}{\sqrt{2}} \\
    \end{array}\right)\!\cdot\!\left(\begin{array}{c}
    \gamma^{\textrm{A}}_{k \uparrow} \\
    \gamma^{\textrm{B}}_{k \uparrow} \\
    \gamma^{\textrm{A}}_{k \downarrow} \\
    \gamma^{\textrm{B}}_{k \downarrow} \\
    \end{array}\right),
\end{eqnarray}
the Hamiltonian becomes $H=\sum_{k} \Upsilon_{k}^{\dagger} \hat{\textrm{T}}^{\dagger} \mathcal{H}(k) \hat{\textrm{T}} \Upsilon_{k}$. Here we use \textrm{A(B)} to denote the Majorana index for the fermionic site $\lambda$ in unit cell $j$: $\gamma^{\textrm{A(B)}}_{k\lambda}=\frac{1}{\sqrt{L}}\sum_{j} e^{-ikj} c^{\textrm{A(B)}}_{j\lambda}$, $l=1,\ldots,L$, $\lambda=1,\ldots,n$ [our notation is slightly different from Kitaev's, and the operators $\gamma^{\textrm{A(B)}}_{k\lambda}$ shall not be confused with the Bogoliubov quasiparticle operators $\gamma_{n\sigma}$ in the main manuscript]. It can be easily checked that matrix $M(k)\equiv -4i\hat{\textrm{T}}^{\dagger} \mathcal{H}(k) \hat{\textrm{T}}$ satisfies the following relation
\begin{equation}
M^{\dagger}(k)=-M(k)=M^{\textrm{T}}(-k),
\end{equation}
which indicates that matrix $M(k)$ is analogous to the matrix $\tilde{B}(p)$ defined above. Therefore the present model we studied in the paper is isomorphic to the Majorana wire considered by Kitaev, but with two fermionic sites (two spin species) in one unit cell \cite{kitaev1}. With this identification, it is not hard to see that our system is in symmetry class D \cite{schnyder}, and the preserved particle-hole symmetry is squared to $+1$ \cite{sato1,sato2}:
\begin{equation}
\hat{\Gamma}\mathcal{H}(k)\hat{\Gamma}^{\dagger}=-\mathcal{H}(-k)^{\ast},~~\hat{\Gamma}=\hat{\Gamma}^{\dagger}=\sigma_{x} \!\otimes\! \mathbb{1}_{2\times2}=\left(\begin{array}{cccc}
    0 & 0 & 1 & 0 \\
    0 & 0 & 0 & 1 \\
    1 & 0 & 0 & 0 \\
    0 & 1 & 0 & 0 \\
    \end{array}\right)\!\!,~~\mbox{and}~~\hat{\Gamma}^{2}=\mathbb{1}_{4\times4}.
\end{equation}

Since our model can be mapped to Kitaev's Majorana wire, we can analogously define the Majorana number $\mathcal{M}(H)$ \cite{kitaev1} as the $\mathbb{Z}_{2}$ invariant to characterize the topological structure of the system. It can be explicitly shown that the topological index defined by the Pfaffian (Pf) based Majorana number is completely equivalent to the Berry phase based $\mathbb{Z}_{2}$ index adopted in the manuscript [see refs.~\cite{ghosh,budich}]. After a lengthy but nontrivial derivation, we can get the following key equation:
\begin{equation}
\mathcal{M}(H)=\textrm{sgn}\!\left\{\textrm{Pf}\!\left[-i\mathcal{H}(k\!=\!0)\hat{\Gamma}\right]\right\}\!\cdot\!\textrm{sgn}\!\left\{\textrm{Pf}\!\left[-i\mathcal{H}(k\!=\!\pi)\hat{\Gamma}\right]\right\}=\frac{\det U(k\!=\!0)}{\det U(k\!=\!\pi)}=(-1)^{\nu}, \label{eq:eqn_key}
\end{equation}
which demonstrates the close connection between different formulations of calculating the $\mathbb{Z}_{2}$ topological invariant in $1$D. $U(k)$ here represents the unitary matrix that diagonalizes the BdG Hamiltonian $\mathcal{H}(k)$. Hence we have the following three equivalent ways to evaluate the $\mathbb{Z}_{2}$ index:
\begin{enumerate}
\item[(1)] Perform the integral over Berry curvature of BdG bands in the first Brillouin zone to find the Berry phase $\nu$, which equals $\frac{i}{\pi}\sum_{E\mbox{(}k\mbox{)}<0}\int^{\pi}_{-\pi} \langle\phi\mbox{(}k\mbox{)}|\partial_{k}\phi\mbox{(}k\mbox{)}\rangle dk$, then we can calculate $(-1)^{\nu}$.
\item[(2)] Diagonalize the BdG Hamiltonian $\mathcal{H}\mbox{(}k\mbox{)}$ at the particle-hole symmetric points $k=0,~\pi$ to construct the unitary matrices $U(k=0,~\pi)$, then the $\mathbb{Z}_{2}$ number is simply the ratio between their determinants.
\item[(3)] Evaluate the Pfaffians of the skew matrices $-i\mathcal{H}(k)\hat{\Gamma}$ at the points $k=0,~\pi$, then the Majorana number $\mathcal{M}(H)$ is the multiplication of their sign functions.
\end{enumerate}
We have checked that all these three methods yield the same results. Because of the length limit of PRL, the above formalism has not been included in the manuscript.

\subsection{III. More details on the topological phase transition}

Firstly, because of the presence of ERD SO interaction, the superfluid order parameter $\Delta_{i}$ is now a complex number, whose real and imaginary parts are both oscillating in the real space, but whose magnitude $\Delta_{q}$ can still be a real and spatially independent constant if we choose the periodic boundary condition without a trap. Secondly, as shown in Eq.~(\ref{eq:eqn_key}), the $\mathbb{Z}_{2}$ topological invariant depends only on the magnitude of the superfluid order parameter $\Delta_{i}$ and its center-of-mass (c.m.) momentum $q$, therefore, although $\Delta_{i}$ changes sign periodically, the topological index is still well defined, which is sufficient to characterize the nontrivial topological structure of the $1$D superfluid system. Finally, because the $1$D model Hamiltonian studied in the manuscript is isomorphic to Kitaev's $1$D superconducting wire, we can also use the Majorana number $\mathcal{M}(H)$ as the subtle indicator of the topological phase transition, but it has been shown above that this number is completely identical to the index $(-1)^{\nu}$ employed in the manuscript. In the following, we will try to present a detailed analysis of the topological phase transition by examining the lowest excitation spectra, and the physical mechanism of the varied FFLO states in the phase diagram of Figure 1(b) [see the manuscript] will be clarified in the next section.

\begin{figure}
\begin{center}
\includegraphics[width=10cm]{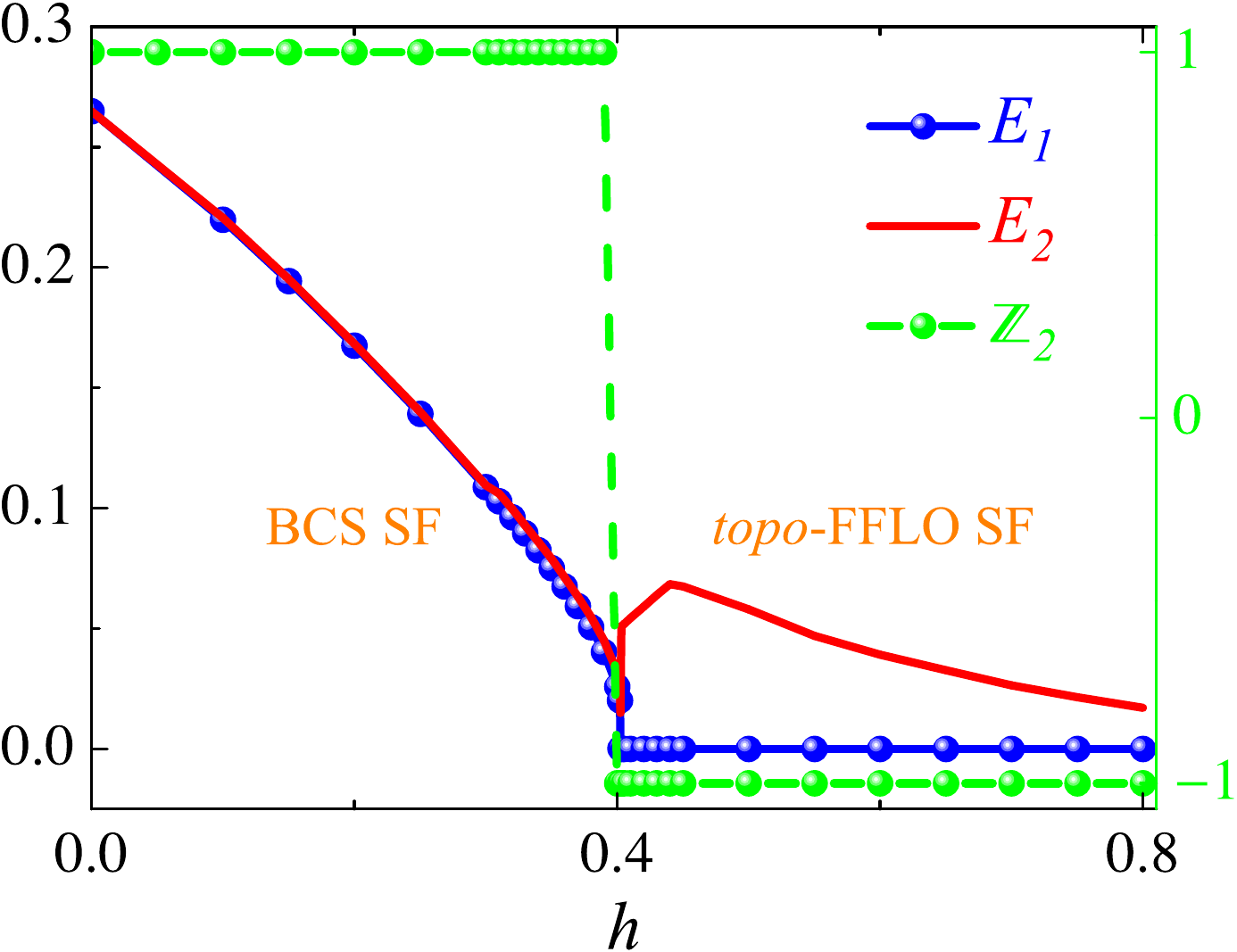}
\caption{\label{fig:reply_fig1} (color online). Two lowest quasihole eigenenergies $E_{1,2}$ as functions of the Zeeman field $h$. The critical point of the topological phase transition between BCS and \emph{topo}-FFLO superfluids can be resolved at $h_{c}\sim 0.4$, where the value of the $\mathbb{Z}_{2}$ invariant changes from $+1$ to $-1$ [see the right $\hat{y}$ axis]. Chemical potential $\mu$ is fixed to be $-1.8$. All the other parameters are the same as in Figure 1(b) of the manuscript.}
\end{center}
\end{figure}

To better understand the revealing topological phase transition in the $1$D ERD SOC Fermi gases, in Fig.~$\ref{fig:reply_fig1}$, we plot the evolution of the two lowest eigenenergies $E_{1,2}$ of the quasihole excitations as increasing the Zeeman field $h$ with fixed chemical potential $\mu=-1.8$. A critical point of the topological phase transition from the conventional BCS superfluid to the newly predicted \emph{topo}-FFLO superfluid can thus be resolved around $h_{c}\sim0.4$, where the calculated $\mathbb{Z}_{2}$ index changes its value from $+1$ to $-1$, indicating the emergence of unpaired zero energy states at the boundaries as $E_{1}$ illustrates. Typically, $E_{2}$ will largely amount to the value of the bulk energy gap, which, as shown in Fig.~$\ref{fig:reply_fig1}$, will first get closed as approaching the critical point $h_{c}$, and then reopen again to protect the resulting edge zero modes so as to ensure the appearance of physically well separated Majorana bound states. The FFLO nature of this topological superfluid phase will be demonstrated in the following section by projecting the pairing term $H_{\Delta}$ into the helicity basis and studying the corresponding dispersion spectra near the Fermi level.

\subsection{IV. Mapping to Kitaev's spinless Majorana wire}

Let us divide the Hamiltonian $H$ into the noninteracting and interacting parts: $H=H_{0}+H_{\textrm{I}}$, $H_{0}=H_{\textrm{K}}+H_{\textrm{R}}+H_{\textrm{D}}$, $\mbox{and}~H_{\textrm{I}}=H_{\Delta}$ [see Equation (1) in the manuscript]. To gain a simple picture of the band structure in the presence of spin-orbit interactions and the Zeeman field, we first consider the effective one-band model without superfluid pairing. The reduced single-particle Hamiltonian $H_{0}$ in the momentum space can be written under the representation $\Psi_{k}'^{\dagger}=\left(\psi^{\dagger}_{k \uparrow}~~\psi^{\dagger}_{k \downarrow}\right)$ as $H_{0}=\sum_{k} \Psi_{k}'^{\dagger} \mathcal{H}'(k) \Psi_{k}'$, where the $2\!\times\!2$ BdG Hamiltonian $\mathcal{H}'(k)$ is given by
\begin{eqnarray}
\mathcal{H}'(k)&=&
    \left(\begin{array}{cc}
    -2t\cos{k}+h-\mu+2\lambda_{y}\sin{k} & 2i\lambda_{z}\sin{k} \\
    -2i\lambda_{z}\sin{k} & -2t\cos{k}-h-\mu-2\lambda_{y}\sin{k} \\
    \end{array}\right),
\end{eqnarray}
which can be readily diagonalized, and the obtained single-particle spectrum is
\begin{equation}
E_{\pm}(k)=-2t\cos{k}-\mu\pm\sqrt{\left(h+2\lambda_{y} \sin{k} \right)^2+4\lambda_{z}^{2}\sin^2{k}}.
\end{equation}
It is revealing to note that with the presence of both Zeeman field $h$ and Dresselhaus SO interaction $\lambda_{y}$, the inversion symmetry of the Bloch bands will be broken. The corresponding spin-mixed helicity basis $\Phi_{k}'^{\dagger}=\left(\phi^{\dagger}_{k,+}~~\phi^{\dagger}_{k,-}\right)$ can be generally expressed as follows:
\begin{eqnarray}
\left(\begin{array}{c}
    \psi_{k \uparrow} \\
    \psi_{k \downarrow} \\
    \end{array}\right)&=&
    \left(\begin{array}{cc}
    u_{k} & v^{\ast}_{k} \\
    -v_{k} & u^{\ast}_{k} \\
    \end{array}\right)\!\cdot\!\left(\begin{array}{c}
    \phi_{k,+} \\
    \phi_{k,-} \\
    \end{array}\right),
\end{eqnarray}
where $\left|u_{k}\right|^{2}+\left|v_{k}\right|^{2}=1$. Then it would be straightforward to show that the interacting Hamiltonian $H_{\textrm{I}}$ contains the designed spinless $p$-wave pairing after being projected into this helical basis \cite{alicea1},
\begin{eqnarray}
H_{\textrm{I}}&=&-\Delta_{q}\sum_{k}\left( \psi^{\dagger}_{\frac{q}{2}+k \uparrow} \psi^{\dagger}_{\frac{q}{2}-k \downarrow} + \textrm{H.c.} \right) \nonumber \\
&=&-\sum_{k}\left(\Delta_{p,+}(k) \phi^{\dagger}_{\frac{q}{2}+k,+} \phi^{\dagger}_{\frac{q}{2}-k,+} + \Delta_{p,-}(k) \phi^{\dagger}_{\frac{q}{2}+k,-} \phi^{\dagger}_{\frac{q}{2}-k,-} + \textrm{H.c.} \right) \nonumber \\
& &-\sum_{k}\left(\Delta_{s}(k) \phi^{\dagger}_{\frac{q}{2}+k,+} \phi^{\dagger}_{\frac{q}{2}-k,-} + \textrm{H.c.} \right),
\end{eqnarray}
where $\Delta_{p,+}(k)=-\Delta_{q}u^{\ast}_{\frac{q}{2}+k}v^{\ast}_{\frac{q}{2}-k}$ and $\Delta_{p,-}(k)=\Delta_{q}u_{\frac{q}{2}-k}v_{\frac{q}{2}+k}$ denote the intraband $p$-wave fermionic pairings with a finite c.m. momentum $q$ inside the upper and lower helicity branches, respectively. $\Delta_{s}(k)=\Delta_{q}\left(u^{\ast}_{\frac{q}{2}+k}u_{\frac{q}{2}-k}+v^{\ast}_{\frac{q}{2}+k}v_{\frac{q}{2}-k}\right)$ denotes the interband $s$-wave pairing. This explicit derivation clearly demonstrates what we mean by the statement in the manuscript that \lq\lq \ldots. It is worth mentioning that to engender a topological superfluid, besides the nontrivial band structure, the system also needs to sustain the channel of an effective spinless $p$-wave pairing. In our model [Eq.~(2)], a nonzero $\Delta_{q}$ will just encode the designed $p$-wave symmetry for the intraband pairing once projected into the helical basis\ldots.'' If the chemical potential $\mu$ is lying within the spin-orbit gap, the only filled band is the lower helicity branch, thus our present model can be isomorphically mapped to Kitaev's spinless Majorana wire, and their mechanisms should be parallel to each other \cite{lutchyn,alicea1,kitaev1}.

\begin{figure}
\begin{center}
\includegraphics[width=9.991cm]{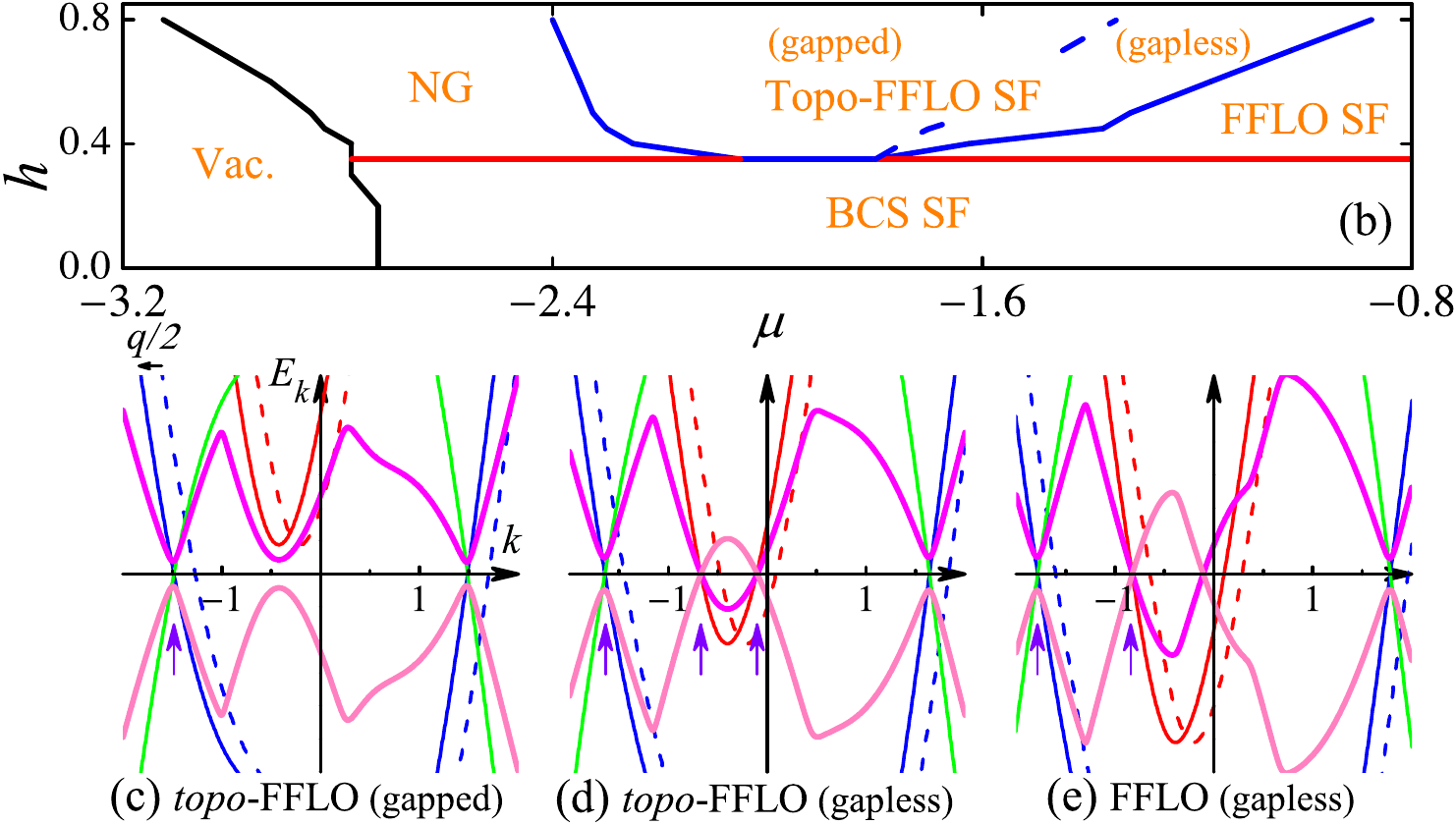}
\caption{\label{fig:reply_fig2} (color online). Panels (b)-(e) from Figure (1) in the main manuscript. Panels (c)-(e) illustrate the dispersion spectra of varied FFLO states in the phase diagram of a $1$D ERD SOC Fermi gas [panel (b)].}
\end{center}
\end{figure}

To clarify the origin of various FFLO states in Fig.~$\ref{fig:reply_fig2}$(b), we highlight here their dispersions near the Fermi level. Figures~$\ref{fig:reply_fig2}$ [(c)-(e)] show the following:
\begin{enumerate}
\item[(1)] Due to the inversion asymmetry, superfluid pairing \emph{inside} the lower \emph{spin-mixed} helicity branch (blue solid line) opens the energy gaps at its Fermi points and shifts the whole bands to the left side by $q/2$. However, since we have forced the system to choose only one specific c.m. momentum $q$, the intraband pairing inside the upper helical band will not be allowed to open extra energy gaps at its inner Fermi points, which gives rise to the gapless state if $\mu$ also crosses the upper helical band (red solid line).
\item[(2)] The $\mathbb{Z}_{2}$ topological invariant can now be approximate to the parity of the number of Fermi points in half of the Brillouin zone (marked by the purple arrows). The inclusion of a finite c.m. momentum $q$ may also generate an additional parameter interval for $\mu$, inside which the noninteracting helicity bands may still have an odd number of Fermi points in half of the Brillouin zone, but three points in one side and one point in the other [see Fig.~$\ref{fig:reply_fig2}$(d)].
\item[(3)] Physically well separated edge modes may appear in the region of gapped \emph{topo}-FFLO phase, namely when the Fermi level lies within the spin-orbit gap. In contrast, the gapless \emph{topo}-FFLO state will typically not possess the stable edge zero modes. However, its inherent topological band structure may still have some nontrivial impacts on its physical properties, especially when the inner Fermi points are also opened, then this parameter region would support Majorana bound states as well. In real systems, the inhomogeneous topological superfluidity should be a broad superposition of fermionic pairings with multiple c.m. momenta, which may help enlarge the gapped regime in the phase diagram.
\end{enumerate}

\subsection{V. Analytical solution of Majorana end state}

To further reveal the properties of zero energy boundary mode between a \emph{topo}-FFLO phase and a non-topological state, we map the model Hamiltonian to a modified $1$D Dirac equation in the low energy (long wavelength) approximation \cite{shen,bernevig}, and analytically establish the connection between our boundary results and the well-known Jackiw-Rebbi solution in one dimension \cite{jackiw}.

First we need to deduce an effective low energy single band model for the \emph{topo}-FFLO state from lattice to continuous space. As is shown, in the helical basis $\Phi_{k}'^{\dagger}=\left(\phi^{\dagger}_{k,+}~~\phi^{\dagger}_{k,-}\right)$ where
\begin{eqnarray}
\left(\begin{array}{c}
    \psi_{k \uparrow} \\
    \psi_{k \downarrow} \\
    \end{array}\right)&=&
    \left(\begin{array}{cc}
    u_{k} & v^{\ast}_{k} \\
    -v_{k} & u^{\ast}_{k} \\
    \end{array}\right)\!\cdot\!\left(\begin{array}{c}
    \phi_{k,+} \\
    \phi_{k,-} \\
    \end{array}\right),
\end{eqnarray}
the total Hamiltonian $H$ [Equation (1) in the manuscript] can be expressed as
\begin{eqnarray}
H&=&\sum_{k} \left\{E_{+}(k) \phi^{\dagger}_{k,+} \phi_{k,+} + E_{-}(k) \phi^{\dagger}_{k,-} \phi_{k,-} -\left(\Delta_{s}(k) \phi^{\dagger}_{\frac{q}{2}+k,+} \phi^{\dagger}_{\frac{q}{2}-k,-} + \textrm{H.c.} \right) \right. \nonumber \\
& &~~~~~~~~~~\left.-\left(\Delta_{p,+}(k) \phi^{\dagger}_{\frac{q}{2}+k,+} \phi^{\dagger}_{\frac{q}{2}-k,+} + \Delta_{p,-}(k) \phi^{\dagger}_{\frac{q}{2}+k,-} \phi^{\dagger}_{\frac{q}{2}-k,-} + \textrm{H.c.} \right)\right\}.
\end{eqnarray}
When the chemical potential $\mu$ is within the spin-orbit gap, we can project out the unfilled upper helicity branch to obtain an effective spinless model,
\begin{equation}
H_{-}=\sum_{k}\left\{ E_{-}(k) \phi^{\dagger}_{k,-} \phi_{k,-}-\left(\Delta_{p,-}(k) \phi^{\dagger}_{\frac{q}{2}+k,-} \phi^{\dagger}_{\frac{q}{2}-k,-} + \textrm{H.c.}\right) \right\},
\end{equation}
where $E_{\pm}(k)=-2t\cos{k}-\mu\pm\sqrt{\left(h+2\lambda_{y} \sin{k} \right)^2+4\lambda_{z}^{2}\sin^2{k}}~\mbox{and}~\Delta_{p,-}(k)=\Delta_{q}u_{\frac{q}{2}-k}v_{\frac{q}{2}+k}$. In the long wavelength limit $\left(|k|\ll1\right)$, we can further do the substitutions that $\left[\sin{k} \rightarrow k\right]$ and $\left[-2\cos{k} \rightarrow (k^{2}-2)\right]$, then the desired continuous spinless model reads
\begin{equation}
H_{-}=\int\frac{dk}{2\pi}\left\{ \left[t_{\textrm{eff}}k^2-2\lambda_y k-\mu_{\textrm{eff}}\right] \phi^{\dagger}_{k,-} \phi_{k,-}-\left(i\frac{\Delta_q \lambda_z}{h}k \phi^{\dagger}_{\frac{q}{2}+k,-} \phi^{\dagger}_{\frac{q}{2}-k,-} + \textrm{H.c.}\right) \right\}, \label{eq:eqn0_pw}
\end{equation}
where $t_{\textrm{eff}} \equiv t-\frac{2\lambda_z}{h}$ and $\mu_{\textrm{eff}}\equiv\mu+2t+h$. We can easily recognize from Eq.~(\ref{eq:eqn0_pw}) that the superfluid pairing has an effective $p$-wave symmetry when $k$ is small, namely $\Delta_{p,-}(k)=-\Delta_{p,-}(-k)$. Written in a new basis $\Phi^{\dagger}_{k,q}=\left(\phi^{\dagger}_{\frac{q}{2}+k,-}~~\phi_{\frac{q}{2}-k,-}\right)$, the reduced Hamiltonian becomes
\begin{eqnarray}
H_{-}&=&\int\frac{dk}{2\pi} \Phi^{\dagger}_{k,q} \mathcal{H}_{-}(k) \Phi_{k,q}, \\
\mathcal{H}_{-}(k)&\equiv&\left(\begin{array}{cc}
    \frac{1}{2}\left[ t_{\textrm{eff}}k^2+m_{\textrm{eff}}+\ell_{q}k\right] & -i\Delta k \\
    i\Delta k & -\frac{1}{2}\left[ t_{\textrm{eff}}k^2+m_{\textrm{eff}}-\ell_{q}k\right] \\
    \end{array}\right).
\end{eqnarray}
Here we define $m_{\textrm{eff}}\equiv\frac{1}{4}t_{\textrm{eff}}q^2-\lambda_{y}q-\mu_{\textrm{eff}}$, $\Delta\equiv\frac{\Delta_q \lambda_{z}}{h}$, and $\ell_{q}\equiv t_{\textrm{eff}}q-2\lambda_y$.

Next let us divide the infinite $1$D chain along $\hat{x}$ direction into two parts:
\begin{enumerate}
\item[(I)] The non-topological region $(x\leq0)$, which is described by a normal $1$D Dirac equation $(\hbar=c=1)$ with a large real positive mass term $m_<$,
\begin{equation}
\mathcal{H}_{<}(k)=k \sigma_x+m_< \sigma_z.
\end{equation}
\item[(II)] The \emph{topo}-FFLO regime $(x\geq0)$, where the low energy effective Hamiltonian $\mathcal{H}_{>}(k)$ is the above-derived $\mathcal{H}_{-}(k)$,
\begin{equation}
\mathcal{H}_{>}(k)=\Delta k \sigma_y+(t_> k^2+m_>) \sigma_z+\ell_> k\mathbb{1}_{2\times2}, \label{eq:eqn1_mde}
\end{equation}
with $t_>\equiv\frac{t_{\textrm{eff}}}{2}$, $m_>\equiv\frac{m_{\textrm{eff}}}{2}$, and $\ell_>\equiv\frac{\ell_{q}}{2}$.
\end{enumerate}
By inspection of Eq.~(\ref{eq:eqn1_mde}), we note that the low energy \emph{topo}-FFLO phase can be described by a modified Dirac equation with two important corrections which are linear and quadratic in momentum, respectively. It shall be emphasized that because positive and negative masses are symmetric in a normal Dirac Hamiltonian, heuristically there is no topological distinction to decide which one is topologically trivial or nontrivial. However, here we follow the literature to assume that Dirac equation with an infinite positive mass term describes the vacuum and is topologically trivial. According to the Jackiw-Rebbi solution of the $1$D Dirac equation, we know that there exists a zero energy bound state in the domain wall between vacua with opposite Dirac masses. Therefore one may speculate that if $m_>$ is negative, a zero mode bound state might appear at the interface $x=0$, which largely amounts to the Majorana end state. Now let us analytically solve the wave function for this bound state solution under small $k$ approximation.

The key spatial differential equation we need to tackle in the region $x\geq0$ is as follows:
\begin{equation}
\left[-i\Delta \partial_x \sigma_y+(-t_> \partial^2_x+m_>) \sigma_z-i\ell_> \partial_x\mathbb{1}_{2\times2}\right] \varphi(x)=0, \label{eq:eqn2_sf}
\end{equation}
where we replace $k$ with the operator $-i\partial_x$ in $\mathcal{H}_{>}(k)$, and assume that the zero energy eigenvector $\varphi(x)$ has the special form
\begin{equation}
\varphi(x)\equiv e^{-\beta x}\left(\begin{array}{c}
    \varphi_1 \\
    \varphi_2 \\
    \end{array}\right), \label{eq:eqn3_wf}
\end{equation}
with $\varphi_{1,2}$ two spatially independent constants. Inserting Eq.~(\ref{eq:eqn3_wf}) into Eq.~(\ref{eq:eqn2_sf}), the resulting secular equation requires
\begin{equation}
\left|\begin{array}{cc}
    -t_{>}\beta^2+m_{>}+i\ell_{>}\beta & \Delta \beta \\
    -\Delta \beta & t_{>}\beta^2-m_{>}+i\ell_{>}\beta \\
    \end{array}\right|=0,
\end{equation}
which gives four solutions of $\beta$,
\begin{equation}
\beta=\pm\sqrt{\frac{\Delta^2+2m_>t_>-\ell^2_>\pm\sqrt{(\Delta^2-\ell^2_>)(\Delta^2+4m_>t_>-\ell^2_>)}}{2t^2_{>}}}. \label{eq:eqn4_beta}
\end{equation}
Eq.~(\ref{eq:eqn4_beta}) can be reformulated in terms of two real positive quantities $\beta_{\textrm{R}}\equiv|\textrm{Re}\beta|$ and $\beta_{\textrm{I}}\equiv|\textrm{Im}\beta|$ as $\beta=\pm \beta_{\textrm{R}}\pm i \beta_{\textrm{I}}$. If $\beta_{\textrm{R}}$ vanishes, the purely imaginary factors $\beta$ will lead to an extended wave function spreading over the half infinite space, which corresponds to a bulk state. Thus we shall demand a finite positive value for $\beta_{\textrm{R}}$ to ensure the existence of a topological phase. Moreover, in order to seek the bound state solution within $x\geq0$, we shall further restrict $\beta=-\beta_{\textrm{R}}\pm i \beta_{\textrm{I}}\equiv\beta_{\pm}$. Once fixing $\beta$, the two constant components of vector $\varphi(x)$ will satisfy the relation
\begin{equation}
\varphi_2=\frac{t_>\beta^2-m_>-i\ell_>\beta}{\Delta \beta} \varphi_1=f(\beta)\varphi_1,
\end{equation}
where $f(\beta)\equiv\frac{t_>\beta^2-m_>-i\ell_>\beta}{\Delta \beta}$. Finally we can obtain the wave function for the zero energy bound state solution of Eq.~(\ref{eq:eqn2_sf}) as a linear combination of the two possible modes:
\begin{equation}
\varphi(x)= e^{-\frac{x}{\xi_>}}\left[e^{i\beta_{\textrm{I}} x}\left(\begin{array}{c}
    \varphi_{+} \\
    f(\beta_{+})\varphi_{+} \\
    \end{array}\right)+e^{-i\beta_{\textrm{I}} x}\left(\begin{array}{c}
    \varphi_{-} \\
    f(\beta_{-})\varphi_{-} \\
    \end{array}\right)\right],~~~x\geq0.
\end{equation}
It can be seen that the wave function $\varphi(x\!\geq\!0)$ has two salient features: (1) This solution is dominantly distributed in the domain wall $x\!=\!0$ and is exponentially decaying into the bulk according to the length scale $\xi_>\equiv \beta^{-1}_{\textrm{R}}$. (2) Owing to the finite imaginary parts of $\beta_{\pm}$, $\varphi(x\!\geq\!0)$ will also display spatial oscillations with the period determined by $\beta_{\textrm{I}}$ before decaying to zero.

In the same way, we can straightforwardly solve the $1$D Dirac equation in the negative-$\hat{x}$ region. The eigenvalue equation is
\begin{equation}
\left(-i\partial_x \sigma_y+m_< \sigma_z\right)\chi(x)=0,~~~x\leq0.
\end{equation}
Assume $m_<$ is a positive Dirac mass, and set the trial wave function
\begin{equation}
\chi(x)\equiv e^{\rho x}\left(\begin{array}{c}
    \chi_1 \\
    \chi_2 \\
    \end{array}\right),
\end{equation}
with $\chi_{1,2}$ two constants, then we can find $\rho=\pm m_<$ through solving the corresponding secular equation. Pick the positive sign and fix the ratio between $\chi_{1,2}$, then the resulting wave function becomes
\begin{equation}
\chi(x)=e^{m_< x}\left(\begin{array}{c}
    \chi_+ \\
    -i\chi_+ \\
    \end{array}\right),~~~x\leq0.
\end{equation}
Now we have three arbitrary constants, which can be determined by imposing the continuity condition at $x\!=\!0$ and the normalization condition, so we would have a unique solution of the bound state. After some algebra, we find our final result
\begin{eqnarray}
\Psi(x)&=&\frac{\mathcal{C}}{\sqrt{2}}\left\{ e^{m_< x}\left(\begin{array}{c}
    1 \\
    -i \\
    \end{array}\right)\left(1-\Theta(x)\right) \right. \nonumber \\
    & &~~~\left.\!+e^{-\frac{x}{\xi_>}}\left[e^{i\beta_{\textrm{I}} x}\left(\begin{array}{c}
    -\frac{i+f(\beta_-)}{f(\beta_{+})-f(\beta_{-})} \\
    -\frac{(i+f(\beta_-))f(\beta_+)}{f(\beta_{+})-f(\beta_{-})} \\
    \end{array}\right)+e^{-i\beta_{\textrm{I}} x}\left(\begin{array}{c}
    \frac{i+f(\beta_+)}{f(\beta_{+})-f(\beta_{-})} \\
    \frac{(i+f(\beta_+))f(\beta_-)}{f(\beta_{+})-f(\beta_{-})} \\
    \end{array}\right)\right]\Theta(x) \right\}, \label{eq:eqn5_bs}
\end{eqnarray}
where $\mathcal{C}$ is a normalization constant, and $\Theta(x)$ is the Heaviside step function with $\Theta(x\!=\!0)\equiv\frac{1}{2}$.

\begin{figure}
\begin{center}
\includegraphics[width=14cm]{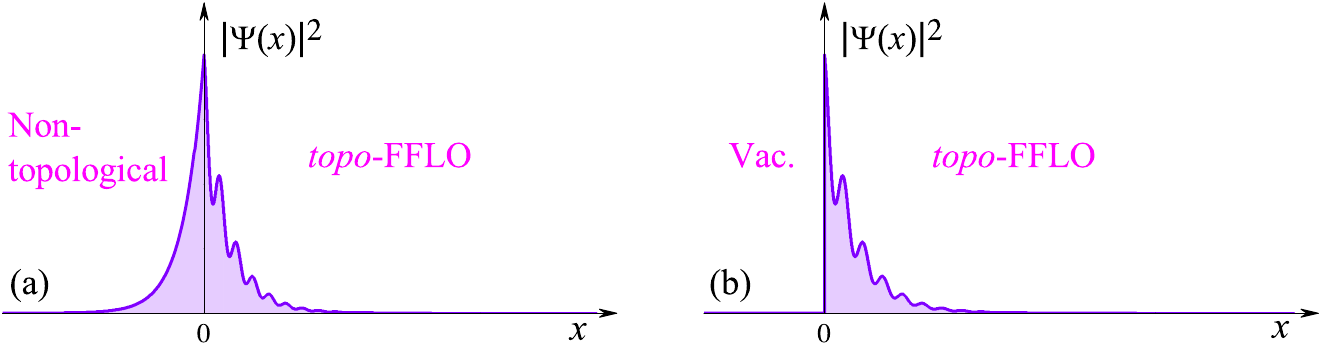}
\caption{\label{fig:fig1_reply_2} (color online). The spatial distribution of the probability density $|\Psi(x)|^2$ for the zero energy bound state Eq.~(\ref{eq:eqn5_bs}). The positive Dirac mass $m_<=+0.5$ in panel (a) and in panel (b), the topologically trivial vacuum has an infinitely large and positive Dirac mass $m_<=+\infty$. The other parameters are fixed to be $t=1$, $\lambda_{z}=\lambda_{y}=0.7$, $h=0.7$, $q=0.45$, $\mu=-1.56$, and $\Delta_{q}=-0.2$ in both cases.}
\end{center}
\end{figure}

In Fig.~$\ref{fig:fig1_reply_2}$, we plot the probability distribution $|\Psi(x)|^2$ of the bound state solution Eq.~(\ref{eq:eqn5_bs}) as a function of the position $x$. Here we have chosen typical values for the parameters: $t=1$, $\lambda_{z}=\lambda_{y}=0.7$, $h=0.7$, $q=0.45$, $\mu=-1.56$, and $\Delta_{q}=-0.2$, which gives rise to a negative Dirac mass in the \emph{topo}-FFLO regime: $m_{>}\simeq-0.74$. While the positive Dirac mass $m_<$ in the non-topological region equals $+0.5$ in panel (a) and $+\infty$ in panel (b), respectively. As manifested by Fig.~$\ref{fig:fig1_reply_2}$, the zero energy eigenmode localized at the boundary $x\!=\!0$ between non-topological and \emph{topo}-FFLO regions resembles the well-known Jackiw-Rebbi solution in the $1$D Dirac equation, whose envelop function exhibits the exponential decay into the bulk systems.

To summarize the above derivations and analyses, we stress the following points:
\begin{enumerate}
\item[(1)] The low energy properties of \emph{topo}-FFLO state can be approximately described by a modified $1$D Dirac equation with both linear and quadratic corrections and a negative Dirac mass.
\item[(2)] If the values of $\beta$ are purely imaginary [see Eq.~(\ref{eq:eqn4_beta})], the system will be topologically trivial, and there will be no zero energy bound mode.
\item[(3)] The length scales of the exponential decay of the bound state equal $m^{-1}_{<}$ and $\beta^{-1}_{\textrm{R}}$ in the non-topological and \emph{topo}-FFLO regions, respectively. While $\beta_{\textrm{I}}$ will determine the period of the oscillatory modifications of the wave function in the topological regime.
\item[(4)] The zero mode boundary state in our system resembles the Jackiw-Rebbi solution and is largely tantamount to the Majorana fermion, which is bound at the domain wall separating regions with opposite Dirac masses or energy gaps.
\item[(5)] Although the analytical solution Eq.~(\ref{eq:eqn5_bs}) captures the main features of the exact numerical results in the main manuscript [see Figs.~3(b),(c),(e),(f)], we shall emphasize that the fully self-consistent calculations in lattice space are more reliable and accurate in detecting and revealing the topological nature of a microscopic model Hamiltonian.
\end{enumerate}
More general comments on the properties of boundary Majorana fermions seem beyond the long wavelength approximation.

\subsection{VI. Comments on the experimental detection of \emph{topo}-FFLO phase}

If the predicted \emph{topo}-FFLO phase can be realized in a real quasi-one-dimensional optical lattice which is constructed from an array of weakly coupled tubes with a large trap aspect ratio \cite{liao}, the most significant physical signals of this quantum state may be detectable at the edges. Experimentally, a direct measurement of the local density of states (LDOS) near the trap edges will provide crucial information pertaining to the associated Majorana fermions and inhomogeneous superfluid pairing of \emph{topo}-FFLO state. In solid state systems, the celebrated scanning tunneling microscope (STM) technique is exactly designed to probe locally the differential conductance, an equivalent of LDOS, of the sample \cite{fischer}. Recent observation of the zero-bias midgap peaks in tunneling spectra of InSb nanowires contacted with the normal metal and superconducting electrodes clearly demonstrates that mapping out LDOS via STM-like powerful tools will give us such valuable information on topological states of matter \cite{mourik}. Therefore, we anticipate that spatially resolved radio-frequency (rf) spectroscopy \cite{shin,schunck}, an analog of STM in cold atom systems, will serve as a suitable technique to detect the described features in LDOS spectra when the condensate enters \emph{topo}-FFLO phase. [In reference \cite{jiang}, Jiang {\it et al.} proposed a modified scheme of implementing the spatially resolved rf spectroscopy to directly yield the LDOS of an ultracold Fermi gas.]

In particular, we can employ the tunability of cold fermions to evolve the system across the boundary of distinct phases, while maintaining the center of the probe at the end so as to resolve visible distinctions in the resulting LDOS spectra that might help differentiate topologically trivial and nontrivial states. For example, we can fix all the other parameters of the system, and just change the value of Zeeman field $h$ to force the condensate into phase transforming from a conventional BCS superfluid to a \emph{topo}-FFLO state, then a remarkable midgap zero-bias peak should appear in the spectrum. Because cold atom systems are intrinsically clean, such zero energy bound states, once showing up, will become unambiguous evidence for the realization of Majorana fermions in the $1$D chain. Previous studies also indicated that this signature of Majorana end state would be stable against the perturbations of intertube tunnelings \cite{mizushima1}. Next we could gradually reduce the strength of Dresselhaus spin-orbit interaction to induce a second phase transition between topological FFLO state and topological BCS state. Typically, the bound midgap peak will be unaffected by such a transition. However, the spectral structure of LDOS near the positive gap edge might probably split into two peaks with reduced weights. Here the main point of our proposal is to identify the topological distinctions of varied phases through tuning the external laser beams to stimulate the evolution of the system across different parts of the phase diagram, and then using the spatially resolved rf spectroscopy to measure the characteristic LDOS spectra. Possible signals consistent with the proposed \emph{topo}-FFLO state might surface out in such means, although finite temperature, trapping potential, fluctuations, and intertube tunneling perturbations may obscure the results of the spectroscopy. Finally, we wish to emphasize that the mechanism of the inhomogeneous pairing in \emph{topo}-FFLO superfluid is quite different from the mechanism of conventional FFLO state, thus even though the ordinary FFLO pairing has eluded definitive observations, it would still be promising to work on this inhomogeneous topological phase, whose detection might bring us double surprises on discovering simultaneously non-Abelian quasiparticles and exotic superfluid/superconducting pairing in one unified and highly controllable setting.

Along with the spatially resolved rf spectroscopy, time-of-flight imaging technique \cite{bloch}, momentum resolved rf spectroscopy \cite{stewart,chen}, and \emph{in situ} density profile measurements \cite{shin2}, as well as measurements of collective modes \cite{nascimbene,edge} could serve as the auxiliary experimental probes to further investigate the rich phase diagram of spin-orbit-coupled Fermi gas systems. Nevertheless, it seems to me that temporarily only spatially resolved rf spectroscopy is capable of detecting the signatures of both Majorana zero modes and the finite center-of-mass momentum pairing in a single measurement.

\subsection{VII. Discussions on the validity of employed mean-field theory}

As being extensively explained that the physical prediction of this new \emph{topo}-FFLO superfluid phase in SOC Fermi gases is a crucial generalization of the recently flourished theory of homogeneous topological superfluidity/superconductivity (which was also emerging from a mean-field theory like ours) to include a nonzero c.m. momentum for the fermionic pairing by explicitly breaking the time reversal and inversion symmetries. Also, the present work demonstrates a novel mechanism for the FFLO superfluidity/superconductivity in SOC systems, which is very different from the conventional FFLO pairing driven by a purely strong Zeeman field. Technically, the exact bosonization and DMRG analyses on such $1$D systems have found that both the FFLO superfluidity and the Majorana bound states are robust and stable against the critical quantum fluctuation corrections \cite{mizushima2,yang1,yang2,stoudenmire,mcheng,fidkowski}, therefore, more or less, confirming and erecting our mean-field predictions on the inhomogeneous topological superfluidity. Moreover, a direct comparison between the mean-field results and the exact Bethe ansatz solutions shows that the BdG theory is useful and reliable for studying the weakly and/or moderately interacting spin-polarized Fermi gases in $1$D \cite{xjliu1}. Hence, as the first endeavor to claim the existence of a new quantum state of matter, our work provides the theoretical framework and presents the basic information of the \emph{topo}-FFLO superfluidity/superconductivity, which shall lay the foundation for future more accurate investigations. Finally, we wish to emphasize that the key content of our work is to predict \emph{for the first time} a new topological state of matter in SOC systems through employing the momentum and real space formulations combined with the appropriate topological arguments.